\documentclass[prd,preprint,nofootinbib]{revtex4-1} 

\usepackage{latexsym,verbatim,enumerate}
\usepackage{amssymb}
\usepackage{amsfonts}
\usepackage{amsmath,color}
\usepackage{graphicx}
\usepackage{bm}
\usepackage[utf8]{inputenc}
\usepackage[T1]{fontenc}
\usepackage[normalem]{ulem}

\usepackage{hyperref}

\hypersetup{
    colorlinks=true,
    linkcolor=red,
    citecolor=green,
    filecolor=magenta,      
    urlcolor=red,
    pdftitle={Low energy limit of stationary axisymmetric dust},
    pdfpagemode=FullScreen,}

\def\ber{\begin{eqnarray}}
\def\eer{\end{eqnarray}}
\def\beq{\begin{equation}}
\def\eeq{\end{equation}}

\def\ed{\end{document}}

\makeatletter
\let\jnl@style=\rm
\def\ref@jnl#1{{\jnl@style#1}}

\def\aj{\ref@jnl{AJ}}                   
\def\actaa{\ref@jnl{Acta Astron.}}      
\def\araa{\ref@jnl{ARA\&A}}             
\def\apj{\ref@jnl{ApJ}}                 
\def\apjl{\ref@jnl{ApJ}}                
\def\apjs{\ref@jnl{ApJS}}               
\def\ao{\ref@jnl{Appl.~Opt.}}           
\def\apss{\ref@jnl{Ap\&SS}}             
\def\aap{\ref@jnl{A\&A}}                
\def\aapr{\ref@jnl{A\&A~Rev.}}          
\def\aaps{\ref@jnl{A\&AS}}              
\def\azh{\ref@jnl{AZh}}                 
\def\baas{\ref@jnl{BAAS}}               
\def\bac{\ref@jnl{Bull. astr. Inst. Czechosl.}}
\def\caa{\ref@jnl{Chinese Astron. Astrophys.}}
\def\cjaa{\ref@jnl{Chinese J. Astron. Astrophys.}}
\def\icarus{\ref@jnl{Icarus}}           
\def\jcap{\ref@jnl{J. Cosmology Astropart. Phys.}}
\def\jrasc{\ref@jnl{JRASC}}             
\def\memras{\ref@jnl{MmRAS}}            
\def\mnras{\ref@jnl{MNRAS}}             
\def\na{\ref@jnl{New A}}                
\def\nar{\ref@jnl{New A Rev.}}          
\def\pra{\ref@jnl{Phys.~Rev.~A}}        
\def\prb{\ref@jnl{Phys.~Rev.~B}}        
\def\prc{\ref@jnl{Phys.~Rev.~C}}        
\def\prd{\ref@jnl{Phys.~Rev.~D}}        
\def\pre{\ref@jnl{Phys.~Rev.~E}}        
\def\prl{\ref@jnl{Phys.~Rev.~Lett.}}    
\def\pasa{\ref@jnl{PASA}}               
\def\pasp{\ref@jnl{PASP}}               
\def\pasj{\ref@jnl{PASJ}}               
\def\rmxaa{\ref@jnl{Rev. Mexicana Astron. Astrofis.}}%
\def\qjras{\ref@jnl{QJRAS}}             
\def\skytel{\ref@jnl{S\&T}}             
\def\solphys{\ref@jnl{Sol.~Phys.}}      
\def\sovast{\ref@jnl{Soviet~Ast.}}      
\def\ssr{\ref@jnl{Space~Sci.~Rev.}}     
\def\zap{\ref@jnl{ZAp}}                 
\def\nat{\ref@jnl{Nature}}              
\def\iaucirc{\ref@jnl{IAU~Circ.}}       
\def\aplett{\ref@jnl{Astrophys.~Lett.}} 
\def\apspr{\ref@jnl{Astrophys.~Space~Phys.~Res.}}
\def\bain{\ref@jnl{Bull.~Astron.~Inst.~Netherlands}}
\def\fcp{\ref@jnl{Fund.~Cosmic~Phys.}}  
\def\gca{\ref@jnl{Geochim.~Cosmochim.~Acta}}   
\def\grl{\ref@jnl{Geophys.~Res.~Lett.}} 
\def\jcp{\ref@jnl{J.~Chem.~Phys.}}      
\def\jgr{\ref@jnl{J.~Geophys.~Res.}}    
\def\jqsrt{\ref@jnl{J.~Quant.~Spec.~Radiat.~Transf.}}
\def\memsai{\ref@jnl{Mem.~Soc.~Astron.~Italiana}}
\def\nphysa{\ref@jnl{Nucl.~Phys.~A}}   
\def\physrep{\ref@jnl{Phys.~Rep.}}   
\def\physscr{\ref@jnl{Phys.~Scr}}   
\def\planss{\ref@jnl{Planet.~Space~Sci.}}   
\def\procspie{\ref@jnl{Proc.~SPIE}}   

\makeatother

\begin{document}


\author{Davide Astesiano}
\email{davide.astesiano@venturilab.ch}
\affiliation{Science Institute, University of Iceland,
Dunhaga 3, 107 , Reykjav\'{\i}k, Iceland}
\affiliation{Mathematics Division, Venturi Space, Route du Pâqui 1, 1720 Corminboeuf, Switzerland}

\author{Matteo Luca Ruggiero}
\email{matteoluca.ruggiero@unito.it}
\affiliation{Dipartimento di Matematica ``G.Peano'', Universit\`a degli studi di Torino, Via Carlo Alberto 10, 10123 Torino, Italy}
\affiliation{INFN - Sezione di Torino , Via Pietro Giuria 1, 1025 Torino, Italy}

\author{Federico Re}
\email{federico.re@unimib.it}
\affiliation{Dipartimento di Fisica \lq\lq Giuseppe Occhialini\rq\rq, Universit\`{a} di Milano Bicocca,\\ Piazza dell'Ateneo Nuovo 1, 20126, Milano, Italy}
\affiliation{INFN - Sezione di Milano Bicocca, Piazza della Scienza 3, 20126 Milano, Italy}

\date{\today}

\title{Non-trivial boundary conditions in general-relativistic models}

\begin{abstract}
We propose an alternative interpretation of dark matter effects within the framework of General Relativity. In particular, we suggest 
 that, in astrophysical and cosmological contexts, different initial assumptions about a system inevitably lead to different interpretations of the same phenomena. As a concrete example, we examine self-gravitating systems composed of an axially symmetric rotating dust fluid and show that effects typically attributed to the presence of additional matter, can instead be reproduced through an appropriate choice of initial and boundary conditions for the equations governing the system.
\end{abstract}

\maketitle

\section{Introduction}\label{sec:intro}
General Relativity (GR) is the most successful theory we currently have for describing gravitational interactions. Its predictions, which have been extensively and accurately tested in a wide range of settings \cite{willrev}, are not merely  corrections to known Newtonian effects, but include genuinely new phenomena with no classical counterpart, such as black holes, gravitational waves and neutron stars.  Remarkably, both strong-field systems near compact objects and weak-gravity systems like the Solar System have self-gravitating configurations of comparable size. Thus, although GR succeeds unequivocally on these scales, its applicability to much larger systems, such as galaxies, is more uncertain. It is difficult  to capture, in just a few lines, the many complications arising from disparate astrophysical and cosmological scenarios and the different levels at which one might seek to describe them. One major reason is that applying GR  at these scales is
complicated by the difficulty of specifying appropriate boundary and initial
conditions: the present state of a galaxy or a cluster of galaxies depends
not only on the matter we currently observe, but also on its entire
formation history and on components that may not be directly visible.
Mathematically, this amounts to setting up a proper Cauchy problem for
Einstein’s equations \cite{MTW}, which dates back to the pioneering work of Lichnerowicz \cite{lichnerowicz1944integration}, Choquet-Bruhat \cite{choquet1962gravitation} and York \cite{york1,york2} and can be summarized as follows:  let $\mathcal{V}_4$ denote spacetime; given a domain
$\Omega \subset \mathcal{V}_4$ and coordinates $x^1,\ldots,x^4$ chosen so
that the condition $g^{44} \neq 0$ is satisfied, we denote by $\Sigma$ the
hypersurface $x^4 = 0$, and we assume that the initial values
$g_{ij}\big|_{\Sigma}$, $g_{ij,4}\big|_{\Sigma}$, $\xi_{(i)}\big|_{\Sigma}$
are prescribed there, where $\xi_{(i)}\big|_{\Sigma}$ are the
thermo-kinetic variables. The goal is to determine the components $g_{ij}$
and the fields $\xi_{(i)}$ inside $\Omega$,  by imposing the validity of
Einstein’s equations. Beyond solving these equations, the major difficulty
lies in how to impose the appropriate conditions on $\Sigma$. All phenomena we aim to describe must therefore be encoded in the degrees of freedom left after fixing the initial conditions. These theoretical difficulties have concrete consequences: cosmological observations appear to require substantial dark energy, while galactic dynamics call for dark matter to explain the data.  \textit{It is evident that the interpretation of phenomena depends sensitively on the initial and boundary conditions, which are unavoidable assumptions that we adopt when we describe such systems.} In this letter, we provide a concrete illustration of this statement, by presenting an alternative, phenomenologically equivalent framework based on different assumptions, and we show how  effects attributed to “dark matter” can be reinterpreted within it. Recognizing the breadth of alternative gravitational theories and how this diversity could  dilute the force of our claim, we  restrict ourselves to GR. The observational manifestations of dark matter span a wide range of physical
scales and dynamical effects (see e.g. \citet{Ciotti:2024uvo} and references therein).  The inferred dark-to-luminous
mass ratio varies significantly among different systems: from spiral (disc) galaxies
to elliptical and dwarf galaxies, and up to galaxy clusters. 
In the generally accepted interpretation, dark matter is invoked to account for several independent classes of phenomena, such as (asymptotically) flat rotation curves in spiral galaxies, the high velocity dispersions of dwarf spheroidal galaxies, gravitational lensing on galactic and cluster scales, the separation between baryonic and total mass inferred in merging clusters (e.g.\ the Bullet
Cluster) and more. It also plays a crucial role in the formation and evolution of galaxies
and large scale structures, as encoded in the standard cosmological model.  Indeed, to date, dark matter is supported exclusively by indirect gravitational evidence, largely interpreted within the Newtonian limit of General Relativity and the $\Lambda$CDM cosmological framework. This circumstance suggests that a deeper understanding of gravitational interactions on large spacetime scales may provide further insight into such observations. Attempting to construct a single, universal set of assumptions for all of these systems, each with its own morphology and evolutionary history, would be naive.  However,
within the framework of GR, without invoking any exotic matter,
we present a line of reasoning that clearly shows how the interpretations are
strongly dependent on the underlying assumptions.

The aim of this letter is twofold: to outline the close link between assumptions and interpretation in astrophysics and cosmology, and to illustrate it through explicit examples. To build a concrete case we necessarily focus on a subset of the relevant phenomena, though in the final section we note how the same techniques can be adapted to other contexts. Namely, we will focus on galaxies, for which the standard theoretical framework depends on what one seeks to compute and at which level of description. However, for the purpose of deriving the leading contribution to the rotational velocities, it is generally accepted that, at the order of approximation considered, time-dependent effects may be neglected (a strong assumption about the initial conditions). The second main assumption is that, if one adopts coordinates adapted to the geometry of the galaxy, the gravitational field, far from the central zone, is expected to be sufficiently weak to be treated as a perturbation of Minkowski spacetime; in this regime, pure general relativistic effects are regarded as negligible, and Newtonian gravity is employed.  \textit{These assumptions (initial and boundary conditions) appear reasonable and well founded, but they are not unique.} We shall show that, even within low-energy GR, one may adopt a different set of assumptions that lead to an alternative interpretation of the same effects we can observe. To provide a concrete analysis, we focus on a relevant class of galaxies that exhibit axial symmetry away from the central bulge region; large families of galaxies approximately satisfy this requirement (for example most disc galaxies). Nonetheless, it will become clear that our line of reasoning can be straightforwardly extended to the different cases mentioned above. 

\section{GR and Galaxies}\label{sec:Galaxies}
As discussed before, when one is interested in computing the leading order
behaviour of the rotational velocities associated with the matter distribution
in a galaxy, the standard treatment is to consider a reference frame adapted to the
approximate symmetry of the system (typically axial symmetry for disc galaxies far from the bulge)
and to assume that the spacetime metric can be written as a weak, static
perturbation of flat Minkowski space. In this regime,
Einstein's field equations reduce to the Poisson equation for the Newtonian gravitational potential, and the geodesic equations reduces to the familiar Newtonian equations of motion. A similar procedure is usually adopted for the evaluation of the gravitational potential responsible for the lensing effects.

It is instructive to contrast this with the cosmological context. As strange as it may sound, if one were to apply the same line of reasoning naively to the Universe as a whole, one might be tempted to conclude that classical Newtonian gravity should suffice to describe its large scale dynamics. However, we know that GR is essential to account for the observed expansion of the Universe, the behaviour of light propagation over cosmological distances, the cosmic microwave background and more.

A crucial point is to understand under which hypotheses the Newtonian theory of gravity can be seen as a limit of GR: \citet{ehlers19} introduced a Frame Theory comprising both the Newtonian theory as well as the Einsteinian one, and showed that in the low-energy limit GR does not coincide with the Newtonian theory, but with a more general one, where an additional ``Coriolis'' field is present: the agreement with GR is obtained imposing suitable boundary conditions.  Here we want to leverage on this difference, which is allowed only if we drop the assumption that the solution is forced to be Newtonian deformation of Minkowski space-time. This is exactly what we do by making use of an important class of physical systems that are consistently described in GR, but whose general low-energy limit is far richer than the corresponding Newtonian description.  We refer to self-gravitating systems composed of an axially symmetric and stationary rotating dust fluid: for these configurations, exact solutions of Einstein's equations are well known and studied \cite{Geroch:1970nt,Geroch:1972yt,HansenWinicour1,HansenWinicour2,Winicour,stephani_kramer_maccallum_hoenselaers_herlt_2003}.  Their peculiarity lies in the fact that compact solutions of these systems are possible in GR only: in fact, no compact or finite dust configuration can exist in Newtonian gravity in the given symmetry conditions \cite{bonnor1977rotating,Ruggiero:2021lpf}, since the only possible solutions is that of an infinite cylinder or that of a thin disc. The fact that such systems can exist in GR  relies on rotation effects which are of the same order of magnitude of Newtonian ones \cite{Ruggiero:2023geh}. In previous works, we studied the low-energy limits of these solutions\cite{Astesiano:2022ozl, Ruggiero:2023geh, Astesiano:2024zzz,Ruggiero:2025tqk}, also in connection with the dynamics of massive particles and photons. 
This class of solutions approximately shares the symmetries of disc galaxies
in regions sufficiently far from the central bulge. To support our claim, it
is enough to consider only a portion $\mathcal{R}$ of a disc galaxy away from its
centre; there is no need to demand that the solution remain valid over the
entire system. Indeed, on galactic scales such a global validity would not be
achievable, owing to the presence of non–axisymmetric features such as bars,
spiral arms, warps, and other perturbations in the mass distribution, which
systematically break exact axial symmetry.

\section{The low-energy limit of the GR solution}\label{sec:lel}


We examine general relativistic solutions describing stationary, axially symmetric configurations of rotating dust, with the aim of demonstrating that differing assumptions can yield significant discrepancies in the interpretation of physical phenomena, even within the same class of solutions. In particular, the solutions in a given spacetime region $\mathcal{R}$ are determined not only by the velocity and density distributions of the matter sources but also, and equally importantly, by the imposed initial and boundary conditions. We focus on their low-energy limit, extensively described in \cite{Astesiano:2022ozl,Ruggiero:2023geh,Astesiano:2024zzz,Galoppo:2024ttc,galoppo1,galoppo2}. To be precise, we define what we mean by ``low-energy limit'': we take the highest order (in powers of $1/c$) consistent system of equations of the general solution, for which in the adapted  reference frame  the 3-velocities $V/c$ and the energy-density $\rho c^2$ are considered small in the usual sense. The solution in the adapted cylindrical coordinates $(x^0 = ct, r, z, \phi)$ is given by
\begin{align}
     ds^2&= -c^2\left(1-\frac{2\Phi}{c^2}-\frac{\psi^2}{r^2c^2}\right)dt^2-2 \psi dt d\phi +  \nonumber \\ & +r^2\left(1+ \frac{2\Phi}{c^2}\right)d\phi^2+ e^{\Psi} \left(dr^2+dz^2\right), \label{eq:metricsolAR} 
\end{align}
and the function $e^{\Psi}$ is determined by the line integrals
\begin{align}
    \partial_{r}\Psi=\frac{1}{2r} \left[2r \partial_r\left(\frac{2\Phi}{c^2}+ \frac{\psi^2}{c^2r^2}\right)+ \frac{(\partial_{z}\psi)^2-(\partial_{r}\psi)^2}{c^2} \right], \nonumber
\end{align}
or, equivalently,
\begin{align}
 \partial_{z}\Psi=\frac{1}{2r} \left[2r \partial_z\left(\frac{2\Phi}{c^2}+ \frac{\psi^2}{c^2r^2}\right)-\frac{2}{c^2} \partial_{r}\psi_{} \partial_{z}\psi_{}\right].  \nonumber
\end{align}
In the above equations, $\Phi$ is equivalent to the usual Newtonian potential\footnote{Here $\Phi$ is defined in analogy with electromagnetism and differs by a minus sign from the standard definition of  the Newtonian potential.}, while the function $\psi$ determines the dragging of the inertial frames \cite{Astesiano:2022ozl}, so we call it \textit{dragging term}; we remark also that, due to the symmetry of the system, all functions depend only on $(r,z)$.  The Newtonian potential $\Phi$ and the dragging term $\psi$ satisfy the following equations:
\begin{eqnarray}
 \rho&=& -\frac{1}{4\pi G} \left[\nabla^2 \Phi+ \frac{(\partial_{z}\psi)^2+(\partial_r\psi-2 \frac{\psi}{r})^2}{2 r^2}\right] \label{Sgmsource2AR} \\
0 &=& \partial_{rr} \psi+ \partial_{zz} \psi- \frac{\partial_r \psi}{r}, \label{Sgmsource1AR} \\
 \frac{V}{r} \partial_r \psi&=&\partial_r \Phi+  \frac{V^2}{r}+ \frac{\psi}{r} \partial_r \frac{\psi}{r},\label{SGMrAR}\\
 \frac{V}{r} \partial_z \psi&=&\partial_z \Phi+ \frac{\psi}{r^2} \partial_z \psi. \label{SGMzAR}
\end{eqnarray}
where $\rho c^2$ is the energy-density of the dust and $V$ is the 3-velocity of dust fluid, as measured by the stationary observer 
We see that $\psi$ is a solution of  Eq.  (\ref{Sgmsource1AR}), which is called Grad-Shafranonv equation \cite{grad1958hydromagnetic,shafranov1958magnetohydrodynamical}: as we remarked,  once the sources are given in the region $\mathcal{R}$, the gravitational potentials $\{\Phi,\psi\}$ are not uniquely determined, since they still depend on the particular solution of this equation which, in turns, requires suitable boundary conditions. By examining  Eqs. (\ref{SGMrAR}) and (\ref{SGMzAR}), which describes the motion of the sources along geodesics with constant $r$ and $z$ coordinates, we observe that the function $\psi$ gives rise to Lorentz-like force terms. In particular,  to guarantee equilibrium in the $z$ direction,  the dragging term $\psi$ should not be negligible compared to $\Phi$: differently speaking, the existence of this system is determined by the rotational effects encoded by $\psi$, and these effects are comparable in magnitude to the Newtonian ones described by $\Phi$ \cite{Ruggiero:2023geh}. According to the field equation (\ref{Sgmsource2AR}), the function $\psi$ modifies the interplay between the sources of the gravitational field and the Newtonian potential, with an effective matter density in the form
\beq
\rho_{\psi}=\frac{1}{4\pi G}\left(\frac{(\partial_{z}\psi)^2+(\partial_r\psi-2 \frac{\psi}{r})^2}{2 r^2} \right).  \label{eq:rhopsi1}
\eeq

Equation (\ref{Sgmsource2AR}) generalizes the Poisson equation to this more general regime, reducing to it when $\psi$ vanishes. As a consequence, a given solution for the Newtonian potential can be referred to an effective energy content which is \textit{greater} than the actual matter content deduced, for instance, in terms of visible matter. 

{
It is important to clarify a crucial point that is often misunderstood. The metric (\ref{eq:metricsolAR}) arises as the low-energy limit of an exact solution of Einstein's equations; however, as discussed in \citet{Astesiano:2024zzz}, this limit does not coincide with the standard post-Newtonian approximation underlying the usual gravitoelectromagnetic analogy \cite{mashhoon03,Ruggiero:2023ker}. In the post-Newtonian regime, the function $\psi$ satisfies
\beq
\partial_{zz} \psi+\partial_{rr} \psi- \frac{\partial_r \psi}{r} = -\frac{16\pi G}{c^2} r\rho V. \label{eq:0GMA}
\eeq
Hence, while in our framework $\psi$ is determined by the homogeneous equation (\ref{Sgmsource1AR}), in the standard post-Newtonian treatment it is directly sourced through the non-homogeneous equation (\ref{eq:0GMA}). As a consequence, within standard gravitomagnetism the dragging terms are typically suppressed by a factor $1/c^2$. By contrast, as discussed above, the existence of our configurations requires rotational effects that are not negligibly small compared with the Newtonian contributions. For this reason, we refer to the limit obtained from the homogeneous equation (\ref{Sgmsource1AR}) as the \textit{strong gravitomagnetic limit}, in order to clearly distinguish it from the conventional post-Newtonian regime.}

\paragraph{The Newtonian approach.} The standard approach to obtain a consistent solution for the system we are considering is based on the assumption that $\psi=0$ in the region $\mathcal{R}$. This condition denies the possibility to have a balance along the $z$ axis unless we adopt the usual procedure which relies on the construction of a razor thin disc solution. As said before, we consider the solution for a region $\mathcal{R}$ far from the galactic center and with surface density $\sigma$. The dynamics in the region is driven by the homogeneous solution ($\nabla^2 \Phi_0(r,z)=0$ in $\mathcal{R}$) for $\Phi_0$, given by the matter $\rho_0$ situated in the inner regions (outside $\mathcal{R}$) and the gravitational potential $\Phi_\sigma$ of the sources $\sigma$: $\Phi_N=\Phi_0+\Phi_\sigma$. Hence, the gravitational potential \(\Phi_N(r,z)\) in $\mathcal R$ satisfies the Poisson equation
\begin{align}
\nabla^2 \Phi_N(r,z) =- 4\pi G \rho(r,z), \qquad
\rho(r,z) = \sigma(r)\,\delta(z),
\end{align}
where \(\sigma(r)\) is the surface density. For \(z \neq 0\), the potential is harmonic and we impose reflection symmetry about the midplane, $\Phi_N(r,z) = \Phi_N(r,-z)$.
The surface density determines the jump of the vertical derivative of the potential at the disc, which in conjunction with the symmetry condition, reduces to
\begin{equation}
\left.\frac{\partial \Phi_\sigma}{\partial z}\right|_{z=0^+}
= -2\pi G\,\sigma(r). \label{eq:jump}
\end{equation}
A convenient representation of the axisymmetric solution of Laplace's equation that satisfies \eqref{eq:jump} is usually obtained via a Hankel (Bessel) transform. The potential may be written as
\begin{align}
\Phi_\sigma(r,z)
&= 2\pi G \int_0^\infty J_0(kr)\,S(k)\,e^{-k|z|}\,{\rm d}k, \\
S(k)
&= \int_0^\infty J_0(kr')\,\sigma(r')\,r'\,{\rm d}r',
\end{align}
 where $S(k)$ is the zeroth-order Hankel transform of the surface density $\sigma(r)$, and $J_0$ is the Bessel function of the first kind. For a given couple $\{\rho_0,\sigma\}$, the equations yield the circular velocity profile 
\begin{equation}
V_N^2(r)
=- r\,\frac{\partial \Phi_N}{\partial r}(r,0).
\end{equation}

Within this Newtonian framework, reproducing an asymptotically flat rotation curve as $r \to \infty$ cannot be achieved by relying solely on vacuum (homogeneous) solutions $\Phi_0$ in the region $\mathcal{R}$, i.e., $\sigma(r) = 0$. In such a vacuum region, the admissible axisymmetric solutions of Laplace’s equation that remain regular at infinity behave as $\Phi_0(r,z) \sim -\frac{GM}{r}$, where $M$ denotes the mass contained in the inner region. Consequently, this behavior entails a Keplerian fall-off of the circular velocity, $V_0(r) \sim r^{-1/2}$. To sustain $V_0(r) \approx \text{const}$ at large radii, one must instead
introduce an additional contribution to the Newtonian potential generated by a
non-vanishing surface density $\sigma$ at large $r$, thereby increasing the total mass
content on the equatorial plane $z=0$.
If we focus on a region $\mathcal{R}$ located well outside the luminous part of
the disc, the visible matter is confined to smaller radii, and its influence
within $\mathcal{R}$ enters essentially through the homogeneous solution for $\Phi_0$ of Laplace's equation in $\mathcal{R}$. The requirement of a non-Keplerian (for instance, flat) rotation curve in this region then forces us, within the Newtonian picture, to
postulate an additional surface density component $\sigma_{\mathrm{DM}}(r)$
supported in $\mathcal{R}$, naturally interpreted as a dark-matter
contribution.The same line of reasoning can be applied for the additional mass required by gravitational lensing. 
\paragraph{A general relativistic alternative.}
Now we compare this result to the case in which we do not assume that $\psi=0$ in the region $\mathcal{R}$. By changing this assumption we will show how the same phenomenology of asymptotic flat velocity (or gravitational lensing) will change. To do the comparison we introduce a deformation of the Newtonian solution just described $(\Phi_0, V_0, \rho_0, \psi=0)$ with no extra matter ($\sigma=0)$ in the region $\mathcal{R}$
\begin{align}
    \Phi= \Phi_0+ \Phi_\psi, \qquad V= V_0+ V_\psi,
\end{align}
by using a nonzero $\psi$. To carry out the deformation, we keep the density fixed, $\rho = \rho_0$, in the inner regions and examine the resulting modifications to the velocity $V$. Therefore, by linearity of the Laplacian, $\Phi_\psi$ and $\psi$ satisfy the following equations
\begin{align}
\partial_{rr} \psi+ \partial_{zz} \psi- \frac{\partial_r \psi}{r}&=0, \label{Vacuum1} \\
    \nabla^2 \Phi_\psi+ \frac{(\partial_{z}\psi)^2+(\partial_r\psi-2 \frac{\psi}{r})^2}{2 r^2}&=0 \label{Vacuum2}.
\end{align}
The additional terms $\{\Phi_\psi,\psi\}$ are not forced to have the same razor thin disc interpretation as the undeformed Newtonian solution.
For our purpose it is enough to choose the following solution
\begin{align}
    \psi(r,z) 
    = C_0\sqrt{r^2+z^2}, \label{eq:defpsi}
\end{align}
where $C_{0}$ is a constant with dimensions of velocity. Substituting in Eq. (\ref{SGMzAR}) we see that $\frac{\partial}{\partial r}\,\Phi_\psi(r,0)=0$ on the equatorial plane.
The function $\Phi_\psi$ has another important property: it has no singularities or jumps (nor do its derivatives) on the plane $z=0$.
Therefore nor $\psi$ nor $\Phi_\psi$ introduces extra matter in the region $\mathcal{R}$.
As in the Newtonian razor–thin disc, we use only the radial component of
the equations of motion to determine the circular velocity on the plane.
Substituting the above solution into the radial velocity equation
\eqref{SGMrAR} we obtain
\begin{align}
    V_{\pm}(r,0)
    =\frac{1}{2}\left(C_0 \pm\sqrt{C_0^2+4 V_0(r)^2}\right),
    \label{Vb-1}
\end{align}
corresponding to co–rotating and counter–rotating solutions, respectively.
In what follows we focus on the co–rotating branch $V(r,0)=V_+(r,0)$.
For large radii the purely Newtonian rotation curve behaves as
$V_0(r,0)^2\sim 1/r$, so $V_0(r,0)\to 0$ and
\begin{align}
    V(r,0)\xrightarrow[r\to\infty]{} C_0 ,
\end{align}
i.e. the deformed rotation curve becomes asymptotically flat and reproduces the observed phenomena at large radii in $\mathcal{R}$. We remark that the limit $r\to\infty$ simply means that we are considering the behaviour in this region, and it is in this region that the expression (\ref{eq:defpsi}) of the function $\psi$ is meaningful. Moreover,
for any $r$ one has $V(r,0)\ge V_0(r,0)$, so the deformation
increases the circular velocity with respect to the Newtonian counterpart with no addition of extra matter in $\mathcal{R}$. In addition, we highlight that the same procedure based on the role of $\psi$ can also be applied to produce Newtonian dark-matter–like effects in gravitational lensing, as shown in \cite{Ruggiero:2025tqk} and \cite{galoppo1}.

\paragraph{A time-dependent scenario.} To further illustrate our approach, we consider the case where
the potentials are allowed to depend on the time coordinate $x^4$ introduced above. The equations corresponding to (\ref{Sgmsource2AR})-(\ref{SGMzAR})  can
be schematically written in the form \begin{align}
    \Box\!\left(\Phi+ \frac{\lvert\vec A\rvert^2}{8}\right) &\sim \rho, \\
    \nabla \wedge (\nabla \wedge \vec A) &\sim \frac{d}{dt}(\nabla \Phi), \\
    \nabla\!\left(\Phi+ \frac{\lvert\vec A\rvert^2}{8}\right)
      +\frac{1}{2} \vec V \wedge (\nabla \wedge \vec A) &\sim \frac{d\vec V}{dt}.
\end{align}
where $\displaystyle \vec A=2\frac{\psi}{r} \hat{\phi}$ and $\vec V=V \hat{\phi}$ and the symbol $\sim$ indicates that we are only displaying some
representative terms, and that the full equations are considerably more
complicated. Proceeding in a schematic manner, but with the same reasoning as in the
previous cases, one could obtain similar effects, now attributing them to
the time dependence of the functions $\Phi$ and $\psi$. In this scenario,
the sources of the effective ``dark matter'' could again lie outside, in
time, the region $\mathcal{R}$: they would be rooted in the past evolution
of these fields. Regarding this, the recent preprint \cite{beordo2025framedraggingvectorpotentialgalaxy} studies the origin of frame-dragging effects from the cosmological past evolution.

In addition, we point out that, because of the interplay between frame-dragging effects and fluid vorticity \cite{Ruggiero:2021lpf}, in a time-dependent scenario such effects may be relevant to dissipative processes occurring during the early phases of galaxy formation, which can generate fluid vorticity in protogalactic clouds.

\section{Discussion}

We have just shown, by means of a concrete example, that when solving Einstein's equations  different choices
of boundary and initial conditions on the initial hypersurface $\Sigma$ can
lead to different interpretations of the same observations. In the first (Newtonian) case we search for sources within the region $\mathcal{R}$; in the second case we are led to look for sources outside $\mathcal{R}$; in
the third case the sources lie outside in time the region $\mathcal{R}$. In particular, our approach has focused on axially symmetric configurations of rotating dust and has shown that, when treated as relativistic systems with appropriate boundary and initial conditions, they exhibit a distinctive non-Newtonian behavior capable of reproducing the observed flatness of rotation curves. In this way, we provide additional theoretical support for recent results suggesting that such configurations can be interpreted as relativistic models of galaxies \cite{crosta2020testing,10.1093/mnras/stae855,galoppo1,Galoppo:2024ttc,Galoppo:2024mfw}. More broadly, the underlying premise of our approach is that, by restricting ourselves to an overly narrow set of assumptions, we risk searching for sources that do not exist or placing them in regions of physical spacetime where they need not reside. As a further remark, we note that observations clearly show that families of galaxies with the same morphology, comparable baryonic content, and similar structural properties can nonetheless exhibit markedly different amounts of dark matter. This is fully consistent with the line of reasoning developed here: such galaxies may have experienced distinct evolutionary histories, characteristic time scales, external environments, and related influences, any of which may place them in different physical states at the time we attempt to describe them.

Indeed, as noted in the Introduction, observations typically attributed to dark matter are diverse and multifaceted: our approach, which may be regarded as a shift in paradigm in the interpretation of dark matter, can be extended, despite the technical challenges posed by the equations, to geometrically more general configurations. As symmetry is reduced, a larger number of functions must be specified, together with explicit assumptions about their boundary behaviour. Even for systems such as the Bullet Cluster, for galaxy clusters more broadly, and for similarly complex structures, the mathematical analysis becomes increasingly intricate; nonetheless, the dependence of the physical interpretation on the underlying assumptions remains the same. This indicates that further insight into the nature of what we call dark matter may arise from applying general relativity, equipped with suitable boundary and initial conditions, to systems whose spatial and temporal scales far exceed those over which Einstein’s theory has been directly tested thus far.

\section*{Acknowledgements}

The authors express gratitude to Dr. Antonello Ortolan for engaging discussions and valuable insights. M.L.R. gives thanks for the support of the Gruppo Nazionale per la Fisica Matematica (GNFM).

\bibliography{refs}

\end{document}